\documentclass[12pt]{article}
\input{amssym.def}
\input{amssym}
\usepackage[dvips]{color}
\usepackage{epsfig}
\usepackage{hyperref}
\normalbaselines
\makeatletter
\@addtoreset{equation}{section}
\makeatother

\def\be{\begin{equation}}
\def\ee{\end{equation}}
\def\ba{\begin{eqnarray}}
\def\ea{\end{eqnarray}}

\newcommand{\rf}[1]{(\ref{#1})}

\def\bra#1{\langle #1|}
\def\ket#1{|#1\rangle}

\begin{document}

\title
{A conformal invariant growth model}

\author{Francisco C. Alcaraz$^1$   and 
Vladimir Rittenberg$^{2,3}$
\\[5mm] {\small\it
$^1$Instituto de F\'{\i}sica de S\~{a}o Carlos, Universidade de S\~{a}o Paulo, Caixa Postal 369, }\\
{\small\it 13560-590, S\~{a}o Carlos, SP, Brazil}\\
{\small\it$^{2}$Physikalisches Institut, Universit\"at Bonn,
  Nussallee 12, 53115 Bonn, Germany}\\
{\small\it$^3$ Department of Mathematics and Statistics, University
of Melbourne, Parkville,}\\ 
{\small\it    Victoria 3010, Australia}
}
\date{\today}
\maketitle
\footnotetext[1]{\tt alcaraz@if.sc.usp.br}
\footnotetext[2]{\tt vladimir@th.physik.uni-bonn.de}

\begin{abstract}
We present a one-parameter extension of the raise and peel 
one-dimensional growth model. The model is defined in the configuration 
space of Dyck (RSOS) paths. Tiles from a rarefied gas hit the interface 
and change its shape. The adsorption rates are local but the 
desorption rates are non-local, they depend not only on the cluster hit 
by the tile but also on the total number of peaks (local maxima) belonging to
all the clusters of the configuration. The domain of the parameter is
determined by the condition that the rates are non-negative. 
 In the finite-size scaling limit, the model is conformal invariant 
in the whole open domain. 
The parameter appears in the sound velocity only. At the boundary of the 
domain, the stationary state is an adsorbing state and conformal invariance is
lost.
 The model allows to check the universality of nonlocal observables in the 
raise and peel model. An example is given.  
\end{abstract}

\section{ Introduction} \label{sect1}

In the study of adsorption-desorption (deposition-evaporation)
processes on a planar surface, the Edwards-Wilkinson \cite{EW} and the
Kardar-Parisi-Zhang \cite{KPZ}  growth models have been extensively researched. The
corresponding dynamical critical exponents $z$ being equal to 2 respectively
3/2. Recently the raise and peel model (RPM) was introduced \cite{GNP}. In
this one parameter dependent model, adsorption is local but desorption is
not (the interface is "peeled"). When the parameter is changed in the
critical domain, $z$ changes too, varying from $z = 1$ to zero. The model was
extended to contain sources at the boundaries \cite{PAV} and defects \cite{DEF}. The RPM
was  also used to check various estimators in information theory \cite{INF}.

  If one fixes the value of the parameter such that $z = 1$, the model has
magical properties. The system has a space-time symmetry (conformal
invariance) and the stationary state probability distribution function
(PDF) has fascinating combinatorial properties \cite{COM} being related to a
two-dimensional equilibrium system \cite{JAN}. This is also the single case where
the system is integrable.

  In the present paper we will consider the RPM at the conformal
invariant point only. With a few exceptions, it is not yet known which
characteristics of the stationary state are universal and the
time-dependent properties of the interface are by and large, uncharted
territory. The aim of this paper is to introduce a new model, dependent on
a parameter $p$, which is in the same universality class as the RPM. This
implies that in the finite-size scaling limit the models coincide for the
whole range of values of $p$. It turns out that in this limit, the single
difference between the models is the value of the sound velocity
$v_s(p)$ which is a function of $p$. Since the models are not local,
finding models belonging to the  same universality class is not an
obvious matter.

  For reasons which will be apparent later on, we call the new model  the
peak adjusted raise and peel model (PARPM), it is presented in Section 2.
Except for $p = 1$, the rates are dependent on the system-size $L$ and on $p$.
The PARPM for $p\neq 1$  is not integrable and therefore all the results we have are based
on Monte Carlo simulations on large lattices. In the stationary states,
the PDF's have no magic properties.

  In Section 3 we show that, in the finite-size scaling limit the density of
contact points is independent of the parameter $p$. This observation is
important since for $p = 1$ one recovers the RPM and for this case, the
density of contact points is known exactly (this knowledge is based on
conformal field theory and combinatorics \cite{PAV}).

  We next examine the time  dependence of the average
number of clusters for various lattices sizes using the Family-Vicsek
analysis \cite{FAV}. The large time 
behavior of the density is given by a few levels of the spectrum of the evolution operator
(Hamiltonian) and it is well understood for $p = 1$. We have observed that by a
rescaling of the time, which implies changing the sound velocity $v_s(p)$ as
a function of $p$, one obtains the same finite-size scaling function of
$tv_s(p)/Lv_s(1)$ for all values of $p$ ($L$ is the system size). In order to
check that this picture is correct, we have done a finite-size scaling
analysis of the spectrum of the Hamiltonians and found the same values for
$v_s(p)$ as the ones determined in obtaining the Family-Vicsek function.

  The sound velocity stays finite for $p$ in the interval $0\leq p < 2$ 
and vanishes
for $p = 2$ (the procedure how to take the limit $p\to 2$ is explained 
in the text).
At $p = 2$ the stationary state is an adsorbing state and conformal
invariance is lost.

  We believe that the results mentioned above are a clear 
demonstration that the PARPM is
in the same universality class as the RPM. These observations opens the
possibility to look for other universal quantities than those 
already considered in the study of the RPM. Since the models are
not local, there is no recipe  to find them. We will show that the average
density of sites where desorption does not take place is universal: for
large values of $L$, its value is independent of $p$.

  Our conclusions are presented in Section 4.

\section{ 
  Description of the peak adjusted raise and peel model}

\begin{figure}
\centering
\includegraphics[angle=0,width=0.5\textwidth] {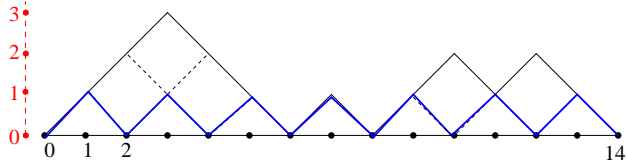}
\caption{
 An example of a Dyck path for L = 14.  There are four contact
points and three clusters. The substrate profile is shown in blue.}
\label{fig1}
\end{figure}

 We consider an open one-dimensional system with $L + 1$
sites ($L$ even). A
Dyck path (restricted solid-on-solid (RSOS) configuration) is defined as
follows. We attach to each site $i$ non-negative integer heights $h_i$
 which
 obey RSOS rules:
\be \label{e4}
 h_{i+1} - h_i =\pm1, \quad h_0 = h_L = 0 \quad  (i = 0,1,\ldots,L-1).
\ee
There are
\be \label{e5}
Z (L) = L!/(L/2)!(L/2 + 1)!
\ee
configurations of this kind. If $h_j = 0$ at the site $j$ one has a {\it
contact point}.
 Between two consecutive contact points one has a {\it cluster}. There are
four contact points and three clusters in Fig.~\ref{fig1}.

%----------------------------------------------------
\begin{figure}
\centering
\includegraphics[angle=0,width=0.5\textwidth] {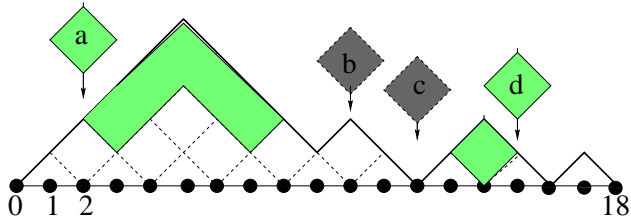}
\caption{
Example of a configuration with 4 peaks of the
PARPM for L = 18. Depending on the position where the tilted tiles reach 
the interface, several distinct processes occur (see the text).}
\label{fig2}
\end{figure}

%---------------------------------------------------
  A Dyck path is seen as an interface separating a film of 
tilted tiles deposited
on a substrate from a rarefied gas of tiles (see Fig.~\ref{fig2}). The stochastic processes in discrete time  has
two steps:
\vspace{0.5cm}

  {\bf a) Sequential updating}. With a probability $P(i)$ a tile hits the site $i =
1,\ldots, L - 1$
($\sum_iP(i) = 1$). In the RPM, $P(i)$ is chosen uniform: $P(i) = P = 1/(L-1)$.
 In the PARPM, this is not longer the case. For a given configuration
(there are $Z(L)$ of them), all the peaks are hit with the same probability
$R_p = p/(L-1)$ ($p$ is a non-negative parameter),  all the other sites are
hit with the same  probability $Q_c = q_c/(L-1)$ where
\be \label{e1}
  q_c = (L-1-pn_c)/(L-1-n_c), \quad      c = 1, 2,\ldots,Z(L).       
\ee
Here $n_c$ is the number of peaks in the configuration (labelled by $c$). $q_c$ depends on the 
configuration $c$ (with $L$ sites and $n_c$ peaks) and on the parameter $p$.
Obviously:
\be \label{e2}
  n_c R_p + (L-1 - n_c)Q_c = 1.          
\ee

\vspace{0.5cm}
  {\bf b) Effects of a hit by a tile}. The consequence of the hit 
on a
configuration is the same as in the RPM at the conformal invariant point.
%----------------------------------------------------
%---------------------------------------------------
Depending of the slope $s_i=(h_{i+1}-h_{i-1})/2$
at the site $i$, the following processes can occur:

\noindent 1) $s_i = 0$ and $h_i > h_{i-1}$ (tile $b$ in Fig~\ref{fig2}). The tile hits a peak and is reflected.

\noindent 2) $s_i = 0$ and $h_i < h_{i-1}$ (tile $c$ in Fig.~\ref{fig2}) . The tile hits a local minimum and is  adsorbed ($h_i\rightarrow  h_i + 2$).

\noindent 3) $s_i = 1$ (tile $a$ in fig.~\ref{fig2}). The tile is reflected after triggering the desorption ($h_j \rightarrow h_j-2$) of a layer of $b-1$ tiles from the segment $\{j=i+1,\ldots,i+b-1\}$ where $h_j>h_i=h_{i+b}$. 

\noindent  4) $s_i = -1$ (tile $d$ in Fig.~\ref{fig2}).  The tile is reflected after
triggering the desorption ($h_j \rightarrow h_j-2$) of a layer of
$b-1$ tiles belonging to the segment $\{j=i-b+1,\ldots,i-1\}$
where $h_j>h_i=h_{i-b}$.

 The continuous time evolution of a system composed by the states
$a = 1,2,\ldots,Z(L)$
with probabilities $P_a(t)$ is given by a master equation that can be interpreted
as an imaginary time Schr\"odinger equation:
\begin{equation}\label{e3}
\frac{d}{dt} P_a(t) = -\sum_b H_{a,b} P_b(t),
\end{equation}
where the Hamiltonian $H$ is an $Z(L)\times Z(L)$ intensity matrix: $H_{a,b}$
non positive ($a \neq b$)
 and $\sum_a H_{a,b} = 0$. $-H_{a,b}$ is the rate for the transition
$\ket b \rightarrow \ket a$. The ground-state wavefunction of the system $\ket0$, $H \ket0 = 0$, gives
the probabilities in the stationary state:
\begin{equation} \label{e4p}
\ket0 = \sum_a P_a \ket a,\;\;\;\;\;\; P_a = \lim_{t \to \infty}  P_a(t). 
\end{equation}

 In order to go from the discrete time description of the stochastic model
to the continuous time limit, we take $\Delta t = 1/(L-1)$ and
\be \label{e5p}
H_{ac}= - r_{ac} q_c \quad (c\neq a),
\ee
where $r_{ac}$ are the rates of the RPM and $q_c$ is given by Eq.~\rf{e1}. The
probabilities $R_p$ don't enter in \rf{e3}  since in the RPM when a tile hits
a peak, the tile is reflected and the configuration stays unchanged.
Notice that through the $q_c$'s the matrix elements of the Hamiltonian
depend on the size of the system and the numbers of peaks $n_c$ of the
configurations.

  As an example, we consider a system with $L = 6$. In this case there 
are 5
configurations shown in Fig.~\ref{figa1}. The Hamiltonian is given by:

%----------------------------------------------------
\begin{figure}
\centering
\includegraphics[angle=0,width=0.5\textwidth] {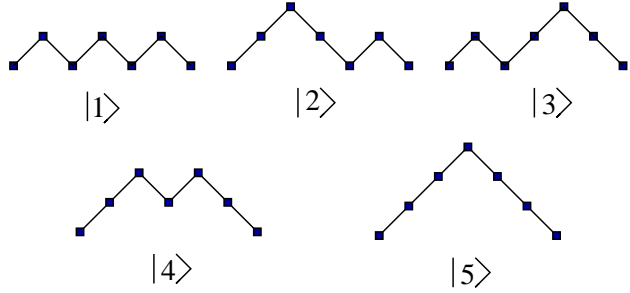}
\caption{\small The five Dyck path configurations for $L=6$}.
\label{figa1}
\end{figure}
%---------------------------------------------------

\ba \label{a3}
&& H =-
\left( \begin{array}{c|rrrrr}
 & \ket{1} & \ket{2} & \ket{3} & \ket{4} &\ket{5} \\ \hline
\bra{1} &   -5 + 3p &    2[(5- 2p)/3]&    2[(5 - 2p)/3] &  0 & 2[(5 - p)/4] \\
\bra{2} &    (5 - 3p)/2 &  -5 + 2p &   0 & (5 - 2p)/3 &      0 \\
\bra{3} &    (5 - 3p)/2 &   0   & -5 + 2p &  (5 - 2p)/3 &    0 \\
\bra{4} & 0  &  (5 - 2p)/3  &  (5 - 2p)/3 &    -5 + 2p &    2[(5 - p)/4] \\
\bra{5} &     0 &  0 &  0 & (5 - 2p)/3 &   -5 + p 
\end{array} \right) .
\ea
  If in \rf{a3} one takes $p = 1$, one recovers the Hamiltonian of the RPM 
for $L =6$  \cite{ARP}.

The unnormalized PDF in the stationary state is given by the eigenvector
corresponding to the eigenvalue zero of (5):

\ba \label{e6}
\left|{\frac{11(5-p)}{2(5-3p)},\frac{15(5-p)}{4(5-2p)}, \frac{15(5-p)}{4(5-2p)},
\frac{3(5-p)}{5-2p},1}\right> .
\ea

Except for $p = 1$ \cite{GNP}, there are no nice combinatorial properties of the
components in \rf{e6} for rational values of $p$. This was checked also for
larger lattices.

  Let us observe that if $p = 5/3$, the configuration $\ket{1}$  which corresponds
to the substrate, becomes an adsorptive state. This phenomenon is
general. For an arbitrary value of $L$, the substrate has $n_p =  L/2$ peaks 
therefore, using \rf{e1} we conclude that for $p = 2(L-1)/L$, the substrate is 
an adsorbing state. For larger values of $p$, one obtains negative rates,
 which  gives
\be \label{e7}
0 \leq p < 2(L-1)/L
\ee
as the domain of $p$.

\section { Conformal invariance in the  PARPM}

  We have invented the PARPM expecting for $p \neq 1$ different properties than
those observed. It came as a surprise that the PARPM 
 belonged to the same
universality class as the RPM. Let us first show that in the stationary
state the density of contact points $g(x,L)$ ($x$ is the distance to the origin
and $L$ the size of the system) is, in the finite-size scaling limit 
($x >> 1$, $L>>1$ but $x/L$ fixed), independent of $p$ and equal to
\be \label{e8}
 g(x,L) = C \left(\frac{L}{\pi}\sin(\pi x/L)\right)^{-1/3},
\ee
where  
\be \label{e9}
C = -\frac{\sqrt 3}{6\pi^{\frac{5}{6}}}\Gamma(-1/6) =
0.753149... \quad .
\ee

\begin{figure}
\centering
\includegraphics[angle=0,width=0.5\textwidth] {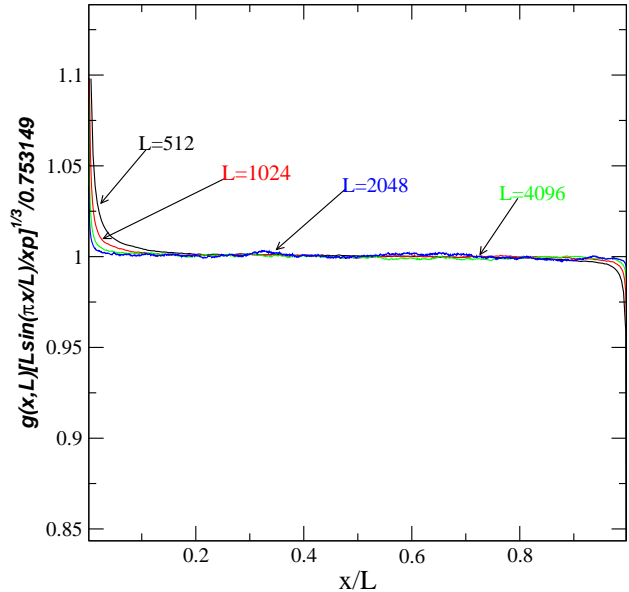}
\caption{
 The density of contact points $g(x,L)$ divided by \rf{e8} for 
$p = 0.01$ 
and lattices sizes $L = 510, 1024, 2048$ and 4096.}
\label{fig3}
\end{figure}

\begin{figure}
\centering
\includegraphics[angle=0,width=0.5\textwidth] {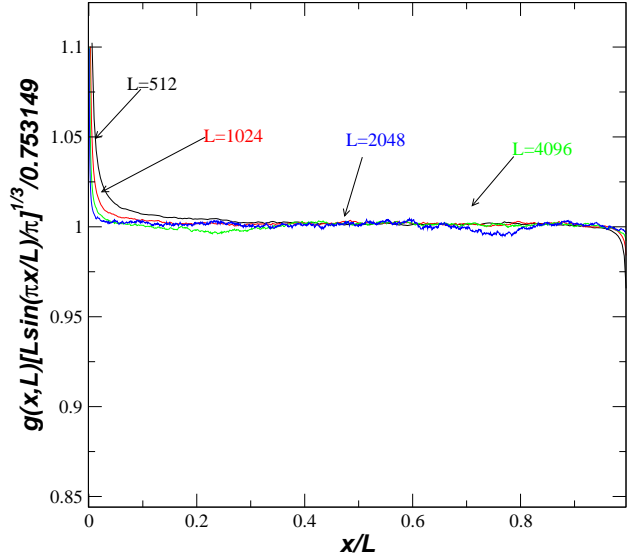}
\caption{
 Same as Fig. 4  for $p = 1.99$.}
\label{fig4}
\end{figure}

  This expression is exact for $p = 1$ \cite{PAV} and its functional form (not the
constant $C$ which is obtained using combinatorics) is a result of conformal
invariance. In Figs.~\ref{fig3}  and \ref{fig4} we show the 
 $x/L$ dependence of the densities of contact points divided by the 
expression \rf{e8}, as
obtained from Monte Carlo simulations for two extreme values of $p$ (0.01
and 1.99)  and various lattices sizes. Similar calculations, done for
other values of $p$, give the same result: in the finite-size limit, the
density of contact points is independent of $p$. From the data collapse
seen in the figures, we conclude that the first test of universality is
successful.

%----------------------------------------------------
%---------------------------------------------------
%----------------------------------------------------
%---------------------------------------------------

 Let us  consider now space-time properties of the model. One starts 
at $t = 0$
with the configuration describing the substrate and we look at the average
number of clusters  $k(t,L)$ as a function of time and lattice sizes. In
order to determine the dynamical critical exponent $z$ for various values 
of $p$,
we compute the Family-Vicsek scaling function \cite{FAV}
\be \label{e10}
K(t,L) = k(t,L)/k(L) - 1,
\ee
where $k(L)$ is the average number of clusters in the stationary state. In 
the finite-size scaling limit one expects
\be \label{e11}
K(t,L) =~ K(t/L^z). 
\ee

%----------------------------------------------------
\begin{figure}
\centering
\includegraphics[angle=0,width=0.5\textwidth] {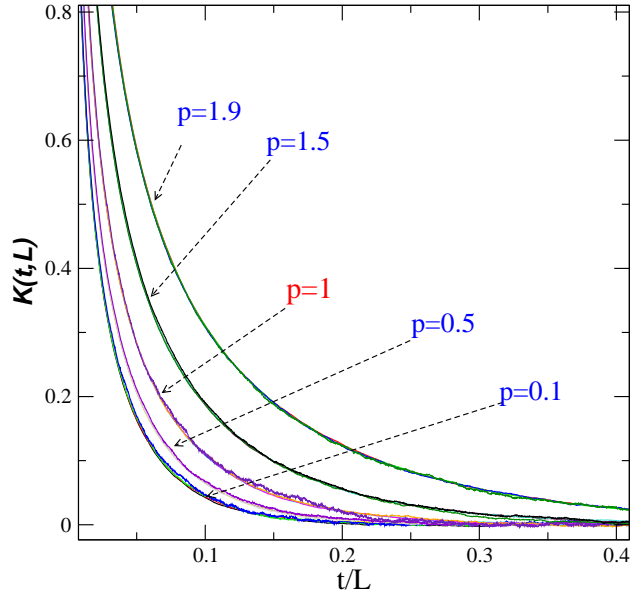}
\caption{
 The Family-Vicsek scaling function $K(t,L)$ as a function of
$t/L$ for $p = 0.1, 0.5, 1, 1.5$ and 1.9 and lattice sizes $L = 1024, 2048,
4096$ and 8192.}
\label{fig5}
\end{figure}
%---------------------------------------------------
  In Fig. \ref{fig5} we show the results of the Monte Carlo simulations. 
For each
value of $p$ one has data collapse which implies $z = 1$ for all the
values of $p$ (see \rf{e11}). One now show that by a change of the time scale,
one can get the different scaling functions in Fig.~\ref{fig5} to coincide. 
Let 
us keep in mind that in a conformal invariant theory, the finite-size
scaling limit of the spectrum of the Hamiltonian  has the following
behavior (the ground-state eigenvalue $E_0$ is equal to zero for a stochastic 
 process):
\be \label{e12}
\lim_{L \to \infty}  E_i(L) = \pi v_s x_i/L,\quad i=0,1,2,\ldots 
\ee 
where $v_s$ is the sound velocity and $x_i$ are critical exponents. This
implies that
\be \label{e13}
K(t/L) = \sum_i C_i \exp(-E_i t) = \sum_i C_i \exp(-\pi v_s x_i t/L),
\ee     
where the constants $C_i$ depend on the initial conditions.
As discussed in \cite{ARP}, for $p = 1$ one has:
\be \label{e14}
v_s(1) = \frac{3}{2} \sqrt{3},\quad   x_1 = 2, \quad x_2 = 3, \quad \ldots \quad .
\ee  
 For large values of $t/L$ one has obtained a very good fit to the data 
using only $x_1$ and $x_2$. A $p$-dependent change of the time scale would
correspond to a change of the sound velocity.

\begin{figure}
\centering
\includegraphics[angle=0,width=0.5\textwidth] {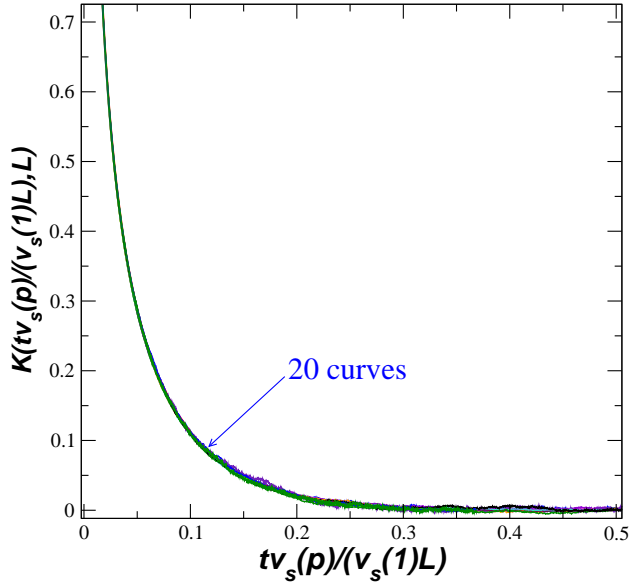}
\caption{
 The Family-Vicsek scaling function $K(tv_s(p),L)$ as a function of
$tv_s(p)/v_s(1)L$ for $p = 0.1, 0.5, 1.5, 1.9$ and $L =1024, 2048, 4096, 8192.$
The 20 sets of data collapse on a single curve.
}
\label{fig6}
\end{figure}

  In Fig. \ref{fig6} we show the data of Fig. \ref{fig5}  choosing for 
$v_s(p)/v_s(1)$ the 
values 1.538 ($p= 0.1$), 1.300 ($p = 0.5$), 0.703 ($p = 1.5$) and 0.472 
($p =1.9$).  Notice that the 20 sets of data collapse on a single curve.

%----------------------------------------------------
%---------------------------------------------------

 In order to check that the determination of the sound velocity is 
correct, we have diagonalized the Hamiltonians up to $L = 26$ sites for 
$p =
0.5, 1$ and 1.5 and looked at the second energy gap $E_2 (p)$. 
The calculation of $E_2(p)$ is easier  to calculate 
numerically as compared with $E_1$ since it corresponds to the 
smallest eigenvalue in the parity odd sector of the 
Hamiltonian.   We have 
computed the ratios $E_2(0.5)/E_2(1)$ and $E_2(1.5)/E_2(1)$ which should
converge for large $L$ to $v_s(0.5)/v_s(1)$ respectively $v_s(1.5)/v_s(1)$.
Extrapolants give the values 1.307 respectively 0.6972 in excellent 
agreement with the values obtained from Family-Vicsek scaling.

  We believe that in this way, we have shown that the PARPM belongs to the
same universality class as the RPM for all values of $p$.

  We give now an example which shows why it is useful to have models in the 
same universality class. Let us consider, in the stationary states, the 
  average density of sites in which the slop vanishes (local minima and maxima):
\be \label{e15}
   \tau(L) = 1/(L-1) \sum_{i=1}^{L-1} (1 - |s_i|).      
\ee
This quantity (ADSSV) was extensively used in the study of the phase diagram of 
the RPM \cite{ARP}. There is a conjecture \cite{YUS} about the values of $\tau(L)$ at the 
conformal invariant point of the RPM:
\be \label{e16}
  \tau(L) = (3L^2 - 2L + 2)/(L - 1)(4L + 2).
\ee   
This conjecture was checked on various lattice sizes. For large values of 
$L$, $\tau(L)$ has the following behavior:
\be \label{e17}
\tau(L) = A + B/L + O(1/L^2),
\ee
where $A = 3/4$ and $B = -1/8$. Are these coefficients universal?
 Using Monte 
Carlo simulations for lattices of different sizes (for $L$ up to 65000) we 
obtained the following results: 
$p =0.5, A= 0.7500, B= 0.219  $; $p = 1, A = 0.7500 , B = -0.125$;  
 $p =1.99, A= 0.7500, B= -0.302$.

 We conclude that the average density of sites where the slope vanishes is 
a universal constant equal to 0.75 with the leading correction term being 
non-universal. An explanation of this interesting result in the context of 
conformal invariance is missing. It is also not clear if the ADSSV is a 
universal observable in other models defined on Dyck paths.

  One can use the asymptotic value of $\tau(L)$ to have an estimate of the 
sound velocity. The average value of the density of peaks, equal to half 
the value of $\tau(L)$, is $<n_c>_{av} = 3/8$. Substituting this value 
in 
\rf{e2}, and using \rf{e5p}  one obtains
\be \label{e18}
<q_c(p)>_{av}  =  \frac{ v_s(p)}{v_s(1)} = 
\frac{8-3p}{5}.
\ee
%----------------------------------------------------
\begin{figure}
\centering
\includegraphics[angle=0,width=0.5\textwidth] {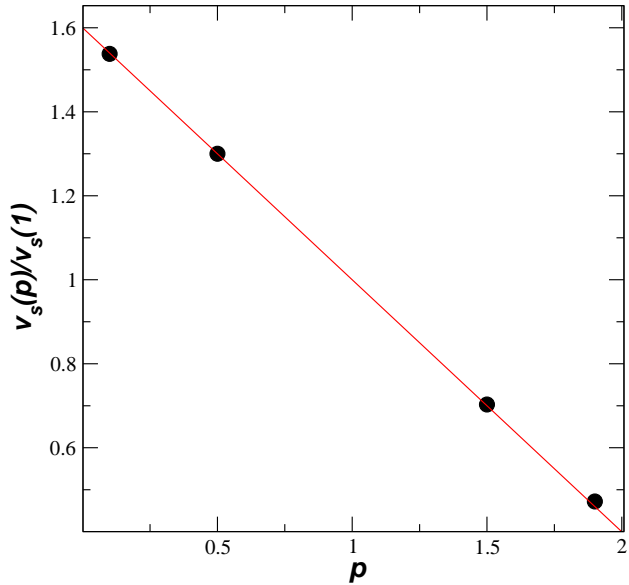}
\caption{
Comparison of the ratios  $v_s(p)/v_s(1)$ given by \rf{e18} (red line) and some of the estimated values (black dots) from the Family-Vicsek  finite-size 
scaling.}
\label{figvs}
\end{figure}
%---------------------------------------------------
We have assumed that the fluctuations of the density of sites where the slop 
vanishes are small.
  One can compare the values obtained for $v_s(p)$ using \rf{e2} with those 
obtained from the diagonalization of the Hamiltonian and from 
Family-Vicsek scaling. In Fig.~\ref{figvs} we compare some of the results 
obtained from the Family-Vicsek scaling with the prediction in \rf{e18}. 
The agreement is excellent.  One 
notices that if $p$ approaches the value $p = 2$, $v_s(p)/v_s(1)$ approaches the 
value 2/5. For the substrate the density of peaks is equal to 1/2, this 
gives, using \rf{e5p} the value $v_s (2) = 0$ for $p = 2$. 
As a consequence, 
$v_s(p)$ decreases from the value 8/5 for $p = 0$ to the value 2/5 if 
$p$ 
 approaches the value 2 from below and has a discontinuity at $p = 2$.
  At $p=2$ the substrate is an absorbing state.

\section{ Conclusions}

 The raise and peel model was up to now the single known conformal 
invariant growth model. The model described in this paper is 
different: it has rates which are dependent on global aspects of the 
configurations (the number of peaks). The fact that conformal invariance is 
maintained came as a surprise. The new model is not integrable and the 
stationary state does not have the remarkable properties of the RPM which 
are a consequence of integrability. Conformal invariance was tested for 
several values of the
 parameter $p$ by computing the density of contact 
points, the spectrum of the evolution operator and the Family-Vicsek 
scaling function. In the finite-size scaling limit, only the sound 
velocity depends on $p$.

 Using the new model we have discovered that the average density of sites 
where the slope vanishes is, for large lattice sizes, a universal quantity and equal to 3/4 for any 
value of $p$. It would be interesting to look for other quantities 
 which are universal.
The model described here can be used as a tool to check universality.

 As explained in the text, the parameter $p$ varies in the domain $0 \leq p < 
2(L - 1)/L$. For each value of $p < 2$, one considers values of $L > 2/(2-p)$ 
which makes sure that the rates are positive. We have observed that 
if $p$ approaches the value 2, the sound velocity decreases 
to a limiting value $v_s=2/5$, and one observes a 
slowing down in the time dependence of physical processes as one 
approaches the stationary state (see Figure 6). 

An interesting phenomenon 
occurs if one chooses $p = 2(L-1)/L$. Conformal invariance is lost and the
stationary state is an absorbing state which is the substrate (see
Figure. 1)). 
Although the stationary state is unique, for lattices sizes 
$L \gtrsim 90$, the relaxation time is very long and 
increases exponentially with the size of the system. One observes 
metastable states. If the initial condition are changed, the system ends up 
in different metastable states. A detailed presentation  of the $p=2(L-1)/L$ case will be published elsewhere \cite{tbp}.

\section{Acknowledgments}

We would like to thank J. Jaimes for discussions and to J. de Gier and P. Pearce for reading the manuscript 
and discussions. This work was supported in part by FAPESP and CNPq 
(Brazilian Agencies), by Deutsche Forschungsgemeinschaft (Germany) and by the 
Australian Research Council (Australia).

\end{document}